\documentstyle[preprint,tighten,aps]{revtex}
\begin{document}
\draft
\date{February 1994}
\title{{\large Strong rescattering in ${K\rightarrow 3\pi}$ decays\\  
and low-energy meson dynamics}\footnote{
Work supported in part by the Human Capital and Mobility Program, EEC
Contract N. CHRX-CT920026.}}
\author{G. D'Ambrosio} 
\address{Istituto Nazionale di Fisica Nucleare, Sezione di Napoli, 
I-80125 Napoli, Italy\\  
Dipartimento di Scienze Fisiche, Universit\`{a} di Napoli, I-80125 Napoli, 
Italy}
\author{G. Isidori, A. Pugliese}
\address{Dipartimento di Fisica, Universit\`a di Roma ``La Sapienza'', 
I-00185 Roma, Italy\\ 
Istituto Nazionale di Fisica Nucleare, Sezione di Roma, I-00185 Roma, 
Italy}
\author{N. Paver} 
\address{Dipartimento di Fisica Teorica, Universit\`{a} di Trieste, 
I-34100 Trieste, Italy\\  
Istituto Nazionale di Fisica Nucleare, Sezione di Trieste, I-34127 Trieste, 
Italy} 
\maketitle
\begin{abstract}
We present a consistent analysis of final state interactions in
${K\rightarrow 3\pi}$ decays in the framework of Chiral Perturbation Theory.
The result is that the kinematical dependence of the rescattering 
phases cannot be neglected. The possibility of extracting the phase 
shifts from future $K_S-K_L$ interference experiments is also analyzed. 
\end{abstract}
\section{Introduction}
\label{sec:intro}
Unitarity requires that final state strong interactions should be taken into 
account in kaon decay amplitudes. On the other hand, such effects are 
interesting by themselves, as they reflect the properties of low-energy 
meson interactions. In this regard, they could represent a meaningful test 
of the theoretical approach based on effective chiral Lagrangians and the 
related amplitude expansions in powers of meson momenta (Chiral Perturbation 
Theory), which should incorporate general features of long-distance 
QCD \cite{leutwyler,georgi}.

The role of ${\pi-\pi}$ phase shifts in ${K\to 2\pi}$ decays has been 
extensively 
discussed \cite{ochs}. These transitions involve the ${\pi-\pi}$ phase 
shifts at 
$\sqrt s=m_K$, and at this energy scale the correction to the leading order 
prediction in Chiral Perturbation Theory (ChPT) are found to be
large in each isospin channel \cite{gasser1}.

In ${K\rightarrow 3\pi}$, final state strong 
interactions act at substantially lower energy 
compared to ${K\to 2\pi}$, and consequently in current fits 
to experimental data 
they have been assumed to be negligible \cite{devlin,kambor}. As a matter of 
fact, these effects could be experimentally accessible in the near future. 

In the approximation of considering only two-body strong interactions, thus 
neglecting the irreducible $3\pi$ rescattering diagrams which should be 
suppressed by phase space, in principle ${K\rightarrow 3\pi}$ 
final state interactions 
can provide a complementary way to study ${\pi-\pi}$ phase 
shifts near threshold. 
Indeed, this is the region where these phase shifts can be most reliably 
predicted in the framework of ChPT. Therefore, such an analysis should 
allow an independent, and significant test of this theoretical approach, to 
be combined with, e.g., those from $K_{l4}$ decays \cite{bijnens}, 
also relevant to the ${\pi-\pi}$ low-energy region. 

As another important point of interest, we recall that the knowledge of 
rescattering is crucial in order to estimate direct CP-violating asymmetries 
in ${K\rightarrow 3\pi}$ \cite{grinstein,isidori1,isidori2,wolf1}. 
Manifestations of such 
asymmetries would allow to determine the existence of direct CP-violation 
in a channel alternative to ${K\to 2\pi}$, and to improve our knowledge of this 
phenomenon, which is predicted by the Standard Model but not clearly 
established yet. 

As emphasized in \cite{dambrosio1}, a convenient way to study final state 
interactions in ${K\rightarrow 3\pi}$ is to consider $K_L-K_S$ 
interference in vacuum as a 
function of time at ``interferometry'' machines such as LEAR \cite{pavlo1} 
and the $\phi$-factory DA$\Phi$NE \cite{paver,fukawa}. The typical 
interference term has the form 
\begin{equation}
{\Re e}\left[\langle 3\pi\vert K_S\rangle^* \langle 3\pi\vert K_L\rangle
\exp{(i\Delta m t)} \right]\exp{(-\frac{\Gamma_S+\Gamma_L}{2}t)},
\label{int1}\end{equation} 
where $\Delta m=m_L-m_S$ is the $K_L-K_S$ mass difference. Indeed, 
studies of the time dependence of Eq. (\ref{int1}) should lead to a 
determination of both the real 
and the (expected small) imaginary parts of the amplitudes. The advantage 
is that the latter ones appear linearly in (\ref{int1}), whereas they appear 
quadratically and consequently are much less accessible in width measurements. 
To obtain an order of magnitude estimate of the effect of the interference 
term, constant phase shifts (independent of energy) were assumed in 
Ref. \cite{dambrosio1}. 

Actually, in the case of ${K\rightarrow 3\pi}$ it is 
not quite appropriate to use the notion 
of constant phase shifts, because: {\it i)} there are two independent $I=1$ 
final states which can be connected by strong interactions, so that one should 
introduce a $2\times 2$ mixing (or rescattering) matrix; and {\it ii)} in 
general the rescattering matrix elements are functions of pions momenta. 

Phenomenologically, ${K\rightarrow 3\pi}$ transition amplitudes for the various 
modes are expanded in powers of the kinematical variables with constant 
coefficients, so that the momentum dependence of rescattering should be 
taken into account for a consistent expansion. This is particularly desirable 
also in connection with momentum expansions predicted by ChPT. Previous 
theoretical estimates of momentum-dependent ${K\rightarrow 3\pi}$ 
strong rescattering were 
performed in the non-relativistic approximation in Ref. \cite{zeldovich} and 
in leading order ChPT for charged kaons in Refs. \cite{grinstein}  
and \cite{isidori1}. In the framework of a complete analysis of all the ChPT 
quartic effects in ${K\rightarrow 3\pi}$, the imaginary parts were 
expanded in the 
kinematical variables and calculated numerically in Ref. \cite{kambor}.

In this paper, we extend the calculation of Refs. \cite{grinstein} and 
\cite{isidori1} to obtain the $K^0\to 3\pi$ rescattering matrix in ChPT. 
To this purpose, we shall first review the general symmetry and unitarity 
constraints on ${K\rightarrow 3\pi}$ amplitudes, then 
we will construct a convenient, 
model independent parametrization of the strong rescattering matrix which is 
unique for all decay modes, and finally we will use ChPT to estimate it. 
The results can be applied to make more reliable predictions for 
the ${K\rightarrow 3\pi}$ time correlations. In addition, our 
analysis allows to clarify 
some delicate questions regarding direct CP-violation in 
$K^{\pm}\rightarrow (3\pi)^{\pm}$.

Specifically, the plan of the paper is as follows: in sect. \ref{sec:forma} 
we set the formalism to expand ${K\rightarrow 3\pi}$ amplitudes; 
in sect. \ref{sec:fsi} 
we define the rescattering matrix and evaluate it in lowest order ChPT; in 
sect. \ref{sec:dalitz} we discuss some consequences for the Dalitz plot 
analysis and CP-violation; in sect. \ref{sec:measurements} we present the 
expectations for the interference term; finally, sect. \ref{sec:concl} 
contains some concluding remarks.

\section{${K\rightarrow 3\pi}$ formalism}
\label{sec:forma}

In the limit of CP conservation, there are five different channels for 
${K\rightarrow 3\pi}$ decays:
\begin{equation} \begin{array}{l}
K^{\pm}\rightarrow \pi^{\pm}\pi^{\pm}\pi^{\mp}  \\
K^{\pm}\rightarrow \pi^{\pm}\pi^{0}\pi^{0}      \\
K_L\rightarrow \pi^{\pm}\pi^{\mp}\pi^{0}        \\       
K_L\rightarrow \pi^{0}\pi^{0}\pi^{0}            \\
K_S\rightarrow \pi^{\pm}\pi^{\mp}\pi^{0}  \end{array} 
\qquad\qquad\begin{array}{l} 
(I=1,2)\\ (I=1,2)\\ (I=1)\\ (I=1)\\ (I=2). 
\end{array}\label{modes}\end{equation}
Here in parentheses we indicate the isospin values relevant 
to the final ($3\pi$) 
states, assuming only  $\Delta I=1/2,3/2$ transitions. In principle, the 
$K_S$ decay to the $I=0$ state is not forbidden, but due to Bose symmetry it is 
strongly suppressed by a high angular momentum barrier \cite{zemach} and we 
neglect it. The first four modes are dominated by $\Delta I=1/2$ transitions, 
while the last one is a pure $\Delta I=3/2$ transition and only recently has 
become accessible through time-dependent 
interference experiments \cite{nakada}. 

For $K(p)\rightarrow \pi_1(p_1)\pi_2(p_2)\pi_3(p_3)$ decays we introduce 
the familiar kinematical invariants 
\begin{equation}
s_i=(p_K-p_i)^2 \qquad{\rm and} 
\qquad s_0={\frac{1}{3}}\sum_i s_i ={\frac{1}{3}} m_K^2 +m_\pi^2,
\end{equation}
where the index $i=3$ refers to the ``odd'' charge pion. Neglecting isospin 
breaking effects, following {\it e.g.} Refs. \cite{zemach,weinberg}, we can 
decompose the decay amplitudes in the general form
\begin{equation} \begin{array}{lll}
A_{++-}&=&2A_c(s_1,s_2,s_3)+B_c(s_1,s_2,s_3)+B_2(s_1,s_2,s_3) \\
A_{+00}&=&A_c(s_1,s_2,s_3)-B_c(s_1,s_2,s_3)+B_2(s_1,s_2,s_3) \\
A^L_{+-0}&=&A_n(s_1,s_2,s_3)-B_n(s_1,s_2,s_3) \\
A^L_{000}&=&3A_n(s_1,s_2,s_3) \\
A^S_{+-0}&=&{\widetilde B}_2(s_1,s_2,s_3). \\ 
\end{array}\label{decomp} \end{equation}
Here, reflecting Bose symmetry and the assumed CP conservation, all 
amplitudes $A_j$ and $B_j$ ($j=c,n,2$) are symmetric under exchange 
$(1\leftrightarrow 2)$. Furthermore, the amplitudes $A_j$ are 
completely symmetric for any permutation of the indices 1,2 and 3.  
Conversely, the amplitudes $B_j$ do not have this symmetry, and under 
permutations of indices only obey the relation 
\begin{equation} 
B_j(s_1,s_2,s_3)+B_j(s_3,s_2,s_1)+B_j(s_1,s_3,s_2)=0.\label{bj}\end{equation}
Finally, the amplitude ${\widetilde B}_2$ is antisymmetric for the 
exchange $(1\leftrightarrow 2)$. It is not independent from the other ones, 
and can be expressed in terms of $B_2$ as
\begin{equation}
{\widetilde B}_2(s_1,s_2,s_3)=
\frac{2}{3}\left[ B_2(s_3,s_2,s_1)-B_2(s_1,s_3,s_2)\right].\label{b2}
\end{equation}
Concerning isotopic spin, the amplitudes $A_j$ and $(B_c,B_n)$ correspond 
to $\Delta I=1/2$ and $\Delta I=3/2$
transitions to the $I=1$ final three-pion state, while $B_2$ is associated 
to the $\Delta I=3/2$ transition to $I=2$.

From the decomposition above we note that there are two amplitudes 
leading to $I=1$ final states, which differ for the 
pion exchange symmetry properties, namely the $A$'s are fully symmetric 
whereas the $B$'s have mixed symmetry. Accordingly, it is convenient to
introduce the two matrices
\begin{equation}
T_c=\left(\begin{array}{rr} 2 &1 \\ 1 &-1 \end{array}\right) 
\qquad
T_n=\left(\begin{array}{rr} 1 &-1 \\ 3 &0 \end{array}\right) ,
\label{tc}\end{equation}
which in the $I=1$ sector transform the symmetric and non-symmetric amplitudes 
into the physical ones for charged and neutral kaons, respectively. Thus: 
\begin{equation}
\left(\begin{array}{c} A_{++-}^{(1)} \\ A_{+00}^{(1)} \end{array}\right)=
T_c \left(\begin{array}{c} A_c(s_i) \\ B_c(s_i) \end{array}\right)
\qquad 
\left(\begin{array}{c} A^L_{+-0} \\ A^L_{000} \end{array}\right)=
T_n \left(\begin{array}{c} A_n(s_i) \\ B_n(s_i) \end{array}\right).
\label{tctn}\end{equation}

Defining the two dimensionless Dalitz plot variables
\begin{equation}
Y=\frac{s_3-s_0}{m_\pi^2} \qquad{\rm and}\qquad X=\frac{s_1-s_2}{m_\pi^2}, 
\label{XY}\end{equation}
and taking into account the symmetry properties of $A$'s and $B$'s, 
we can expand the five independent amplitudes in Eq. (\ref{decomp}) 
in powers of $X$ and $Y$ up to quadratic terms:
\begin{eqnarray}
&A_j=&a_j+c_j(Y^2+X^2/3) \nonumber \\
&B_j=&b_jY+d_j(Y^2-X^2/3).\label{ajbjcjdj}
\end{eqnarray}
Substituting Eq. (\ref{ajbjcjdj}) in Eq. (\ref{decomp}), we obtain 
\begin{equation} \begin{array}{lll}
A_{++-}&=&2a_c+(b_c+b_2)Y+2c_c(Y^2+X^2/3)
	+(d_c+d_2)(Y^2-X^2/3) \\
A_{+00}&=&a_c-(b_c-b_2)Y+c_c(Y^2+X^2/3)
	-(d_c-d_2)(Y^2-X^2/3) \\
A^L_{+-0}&=&a_n-b_nY+c_n(Y^2+X^2/3)
	-d_n(Y^2-X^2/3) \\
A^L_{000}&=&3a_n+3c_n(Y^2+X^2/3)\\
A^S_{+-0}&=&\frac{2}{3}b_2X-\frac{4}{3}d_2XY.
\label{expan11} \end{array} \end{equation}

This decomposition can be easily related to the one introduced in 
Refs. \cite{devlin,kambor}.

\section{$3\pi$ final state interaction}
\label{sec:fsi}

Since strong interactions are expected to mix the two $I=1$ 
final states, we must introduce a strong interaction rescattering matrix  
which mixes the corresponding decay amplitudes. Projecting the final 
state $(3\pi)_{I=1}$ by means of the matrices $T_c^{-1}$ and $T_n^{-1}$ in the 
symmetric-nonsymmetric basis, we can define the scattering 
matrix $R$, common to charged and neutral channels, as follows:
\begin{equation}
\left(\begin{array}{c} A_{++-}^{(1)} \\ A_{+00}^{(1)} \end{array}\right)_R=
T_c R \left(\begin{array}{c} A_c \\ B_c \end{array}\right)
= T_c R T_c^{-1} \left (\begin{array}{c} A_{++-}^{(1)} \\ A_{+00}^{(1)} 
\end{array}\right) \label{scatt1}
\end{equation}
\begin{equation}
\left(\begin{array}{c} A^L_{+-0} \\ A^L_{000} \end{array}\right)_R=
T_n R \left(\begin{array}{c} A_n \\ B_n \end{array}\right)
= T_n R T_n^{-1} \left(\begin{array}{c} A^L_{+-0} \\ A^L_{000} 
\end{array}\right).
\label{scatt2} \end{equation}
Here the subscript $R$ means that in the decay amplitude rescattering has 
been included. The matrix $R$ defined above has diagonal elements which 
preserve the symmetry properties under pion exchanges, as well as 
off-diagonal elements which connect symmetric amplitudes to non-symmetric 
ones and {\it viceversa}.

The unitarity conditions are obtained by imposing conservation 
of probability, namely:
\begin{eqnarray}
\int d\Phi\left[\hskip 2pt \vert (A_{++-}^{(1)} )_R \vert^2 + 
	\vert (A_{+00}^{(1)} )_R \vert^2 \hskip 2pt\right]&=&
\int d\Phi\left[\hskip 2pt \vert A_{++-}^{(1)}  \vert^2 + 
	\vert A_{+00}^{(1)}\vert^2\hskip 2pt\right] \label{unit1},\\  
\int d\Phi\left[\hskip 2pt \vert (A^L_{+-0})_R \vert^2 + 
	\vert (A^L_{000})_R \vert^2 \hskip 2pt\right]&=&
\int d\Phi\left[\hskip 2pt \vert A^L_{+-0}\vert^2 + 
	\vert A^L_{000}\vert^2\hskip 2pt \right],\label{unit2} 
\end{eqnarray}
where $d\Phi$ represents the phase space element. 

We now perform the calculation of $R$ using ChPT. At the lowest order $p^2$, 
there are no quadratic terms and the coefficients $a_j$, $b_j$ in 
(\ref{ajbjcjdj}) are real if CP is conserved. At order $p^4$ loops and 
counterterms will appear, generating real parts with higher powers in $X$ and 
$Y$ and also imaginary parts proportional to the O($p^2$) constants $a_j$ 
and $b_j$ . These imaginary parts define the rescattering matrix relevant to 
the constant and linear terms. In principle, imaginary parts of quadratic 
terms can occur similarly, but since these ones appear at the higher order 
$p^6$ we neglect them, as it is also justified by the smallness of the 
experimental values of quadratic slopes. In this approximation, we replace 
$A_{c,n}$ and $B_{c,n}$ in (\ref{scatt1}) and (\ref{scatt2}) by $a_{c,n}$ and 
$b_{c,n}Y$ respectively. The unitarity conditions resulting from 
(\ref{unit1}) and (\ref{unit2}) are equivalent, and take the form   
\begin{eqnarray}
\int d\Phi\left[ \vert R_{11}\vert^2 +\frac{2}{5} 
\vert R_{21}\vert^2\right]&=&
\int d\Phi \\  
\int d\Phi\left[\vert R_{22}\vert^2 +\frac{5}{2}
\vert R_{12}\vert^2\right]Y^2&=&
\int d\Phi\hskip 2pt Y^2 \\ 
\int d\Phi\left[ 5R_{11}R_{12}^*+2R_{21}R_{22}^*\right]Y&=&0.\label{unit3} 
\end{eqnarray}

As anticipated, at O($p^2$) there are only tree diagrams  
that can be easily computed 
with the leading order chiral weak Lagrangian \cite{cronin}, and 
there is no final state interaction so that $R=I$ (trivial case). At order 
$p^4$ loop diagrams (Fig.1) generate 
imaginary parts, corresponding to on-shell propagators 
in internal lines, so that we can write
\begin{equation}
R=I+i\left( \begin{array}{cc} \alpha(s_i) &\beta'(s_i) \\
\alpha'(s_i) & \beta(s_i) \end{array}\right).
\end{equation}
Using the strong  $O(p^2)$ chiral Lagrangian
\begin{equation}
{\cal L}_S^{(2)} = \frac{F_\pi ^2}{4} {\rm tr} \left[\partial_\mu \Sigma 
\partial^\mu \Sigma^\dagger +M(\Sigma+\Sigma^\dagger) \right],\label{lagra}
\end{equation}
where $\Sigma={\rm exp}(i\phi/ \sqrt{2}F_\pi)$, $\phi$ is the octect
matrix of the pseudoscalar fields, $F_\pi$ is the pion decay constant
($F_\pi \simeq 93$ MeV) and $M={\rm diag}(m^2_\pi, m^2_\pi, 2m_K^2-m_\pi^2)$,
we find
\begin{eqnarray}
\alpha(s_i)&=&\frac{1}{32\pi F_\pi^2}\sum_{i=1}^3 \frac{1}{3}v_i(2s_i+m_\pi^2),
  \label{alpha}\\
\alpha'(s_i)&=&\frac{1}{32\pi F_\pi^2} \sum_{i=1}^2 \frac{5}{3}
  \left[v_i(s_i-m_\pi^2)-v_3(s_3-m_\pi^2)\right],\label{alpha'}\\
\beta(s_i)&=&\frac{1}{32\pi F_\pi^2}\left[\hskip 2pt \frac{1}{3}\sum_{i=1}^3 
  v_i(s_i-4m_\pi^2) +\sum_{i=1}^2 m_\pi^2 \frac{v_3(s_3-s_0)
  -v_i(s_i-s_0)}{s_3-s_0 }\hskip 2pt \right], \label{beta}\\
\beta'(s_i)&=&\frac{1}{32\pi F_\pi^2}\sum_{i=1}^3\frac{2}{3} v_i
  \frac{(s_i-m_\pi^2)(s_0-s_i)}{s_3-s_0},\label{beta'}
\end{eqnarray}
where $v_i$ are the ``velocities'': $v_i=(1-4m_\pi^2/s_i)^{1/2}$.
At this order, only the unitarity condition (\ref{unit3}) is nontrivial
 and implies 
\begin{equation}
\int d\Phi \left[ 2\alpha'(s_i)-5\beta'(s_i)\right]Y =0.
\end{equation}
This condition is exactly verified by the functions in Eqs. (\ref{alpha'}) and 
Eq. (\ref{beta'}), as expected since ChPT is an effective field theory 
where unitarity is perturbatively satisfied.

We remark that the rescattering matrix $R$ could have been directly evaluated 
by just integrating the $\pi-\pi$ scattering amplitude over the phase space 
of intermediate particles. Actually, once  the matrix $R$ has been defined, 
one could improve the lowest order Eqs. (\ref{alpha})-(\ref{beta'}) by 
including all higher orders in strong interactions, or even by replacing them 
by any other available phenomenological information on ${\pi-\pi}$ scattering.  
Analogously, for the weak amplitudes $a_c,b_c$ and $a_n,b_n$ one
could use either the ChPT predictions or the available experimental 
determinations. 

For the decay into the $I=2$ final states there is only one amplitude, with 
definite symmetry under pion exchange which must be preserved by strong 
interactions. Thus, we can write
\begin{eqnarray}
(B_2)_R&=&b_2Y(1+i\delta(s_i))  \nonumber \\
({\widetilde B}_2)_R&=&\frac{2}{3}b_2X(1+i{\widetilde \delta(s_i)}), 
\label{scatt3} \end{eqnarray}
where two functions $\delta$ and ${\widetilde \delta}$ are 
again not independent because $(B_2)_R$ and $({\widetilde B}_2)_R$
must satisfy Eq. (\ref{b2}). At the lowest non trivial order in ChPT we 
find
\begin{eqnarray}
\delta(s_i)&=&{\frac{1}{32\pi F_\pi^2}}\hskip 2pt\bigg[\hskip 1pt
{\frac{1}{3}}\hskip 1pt \sum_{i=1}^3 v_i(s_i-4m_\pi^2) \nonumber\\
&+&{\frac{1}{3}}\hskip 1pt \sum_{i=1}^2{\frac{v_i(s_i-s_0)(2s_i-5m_\pi^2)
 -v_3(s_3-s_0)(2s_3-5m_\pi^2)}{s_3-s_0}}\hskip 1pt\bigg]. 
\label{delta}\end{eqnarray} \par
Regarding three-body rescattering, which will appear at two loops, we would 
expect its contribution to the functions $\alpha$, $\alpha'$, $\beta$, 
$\beta'$ and $\delta$ to be rather small, as being suppressed 
by phase space, assuming that the three-body coupling is not anomalously 
large. This is indeed the case for the leading order Lagrangian 
(\ref{lagra}). Regarding higher orders in ChPT, $O(p^4)$ contributions to 
$R$ might be relevant, similar to the case of ${\pi-\pi}$ 
phase shifts where the 
scattering lengths turn out to be affected at the $20\%-30\%$ level 
\cite{gasser1}. We can take these figures as an indication for the accuracy 
of the subsequent applications of Eqs. (\ref{alpha})-(\ref{delta}).   

\section{Consequences for Dalitz Plot analysis and CP-violation}
\label{sec:dalitz}
As expected from the smallness of the available phase space, the functions 
$\alpha$, $\alpha'$, $\beta$, $\beta'$ and $\delta$ are smaller than unity 
over the whole Dalitz plot. Indeed, by expanding in powers of $X$ and $Y$ 
up to quadratic terms, we obtain
\begin{equation} \begin{array}{llllll}
\alpha(X,Y)&\cong&\alpha_0 + \alpha_1 (Y^2+X^2/3) &[\ \alpha_0\simeq 0.13 
         &\ \alpha_1\simeq -2.9\times 10^{-3} &]\\ 
\alpha'(X,Y)&\cong&\alpha'_0Y + \alpha'_1 (Y^2-X^2/3) &[\ \alpha'_0\simeq -0.12 
         &\ \alpha'_1\simeq 3.4\times 10^{-3} &]\\
\beta(X,Y)&\cong&\beta_0 + \beta_1 (Y^2-X^2/3)/Y \qquad &[\ \beta_0\simeq 0.047
         &\ \beta_1\simeq 4.7\times 10^{-3}   &]\\
\beta'(X,Y)&\cong&\beta'_0(Y^2+X^2/3)/Y  &[\ \beta'_0=\alpha'_0/5 & &] \\ 
\delta(X,Y)&\cong&\delta_0 + \delta_1 (Y^2-X^2/3)/Y  &[\ \delta_0= -\beta_0 
         &\ \delta_1\simeq -0.020 &], 
\end{array}\label{expan}\end{equation}
Using Eqs. (\ref{scatt1}), (\ref{scatt2}), (\ref{scatt3}) 
and (\ref{expan}) we can expand both real and imaginary parts of all 
${K\rightarrow 3\pi}$ amplitudes up to linear terms. 

As a first application of the formalism we can discuss 
the role of rescattering in CP-odd charge asymmetries in $K^\pm\to 3\pi$. 
Contrary to ${K\to 2\pi}$, where direct CP-violation is 
suppressed by the smallness 
of the $\Delta I=3/2$ amplitude, in ${K\rightarrow 3\pi}$ an 
observable effect can  
potentially arise also from the interference of the two $\Delta I=1/2$ 
amplitudes. For a nonvanishing effect it is crucial that the relevant 
amplitudes have different electroweak phases (which can be the case 
only at order $p^4$ in the framework of ChPT) as well as different 
rescattering phases. 

Let us consider, for example, the amplitudes for 
$K^+\rightarrow (\pi^+\pi^+\pi^-)_{I=1}$. From the preceding relations we 
easily obtain 
\begin{eqnarray}
{\Re e} (A_{++-}^{(1)})&=&2a_c+b_cY \\ 
{\Im m} (A_{++-}^{(1)})&=&2a_c\alpha_0+a_c\alpha'_0Y+b_c\beta_0Y=
2a_c\alpha_0+b_cY\left(\beta_0+\frac{a_c}{b_c}\alpha'_0\right)\label{exp1}
\end{eqnarray}
In Eq. (\ref{exp1}) the contribution of $\alpha'_0$ is multiplied by the 
sizable factor $\vert a_c/b_c\vert \simeq 3.5-4.0$, and dominates over the 
one of $\beta_0$ by almost one order of magnitude. This shows that the 
kinematical dependence of rescattering functions is relevant in constructing 
the imaginary parts of the amplitudes. Nevertheless, such a large imaginary 
contribution to the term linear in $Y$ does not help in generating the 
large CP-violating interference between the two $I=1$ amplitudes suggested in 
Ref. \cite{belkov}. Indeed, of the two $Y$-dependent terms in 
Eq. (\ref{exp1}), the one proportional to $a_c$ has the same weak phase as 
the constant term and consequently, as already noticed in 
Ref. \cite{isidori1,isidori2}, the CP-violating interference between the 
amplitudes to the two $I=1$ states must be proportional to the small 
difference $(\alpha_0-\beta_0)$. 

In principle, the rescattering phases should be included in the 
analysis of CP conserving Dalitz plot parameters. Their contribution 
could affect the determination of the linear and the quadratic slopes. 
However, since in this case the imaginary parts appear quadratically, their 
effect is of order $p^8$ in ChPT and thus for completeness also the other
contributions of the same order should be included. As a curiosity, we estimate 
the contribution of 
${\Im m} A^L_{000}$ to the quadratic slope $h$, in the Dalitz plot of 
$K_L\to 3\pi^0$, defined by
\begin{equation}
\vert A^L_{000}\vert^2 = 
({\Re e} A^L_{000})^2+({\Im m} A^L_{000})^2 \propto 1+h(Y^2+X^2/3)+...
\label{hslope}\end{equation}
Using Eqs. (\ref{scatt2}) and (\ref{expan}), we find:
\begin{equation}
{\Im m} A^L_{000}=3a_n\alpha_0+3(b_n\beta'_0+a_n\alpha_1)(Y^2+X^2/3),
\end{equation}
which gives the contribution to $h$\footnote{This numerical result is 
obtained with the value of $b_n/a_n$ resulting from the fit of 
Ref. \cite{kambor}.}
\begin{equation}
h^{\rm (Im)} = 2\alpha_0\left(\alpha_1+ \frac{b_n}{a_n}\beta'_0\right)
\simeq +1.4\times 10^{-3}. \label{hnumval}
\end{equation}
This number turns out to be of the same order of the experimental
value of $h$\cite{somalwar},
\begin{equation}
h = -(3.3\pm 1.1)\times 10^{-3},
\label{hexp}\end{equation}
but is substantially smaller than the $p^6$ contribution theoretically 
estimated in Ref. \cite{donoghue}. 

\section{Measurements of the rescattering matrix in interferometry machines}
\label{sec:measurements}
As pointed out in sect. \ref{sec:intro}, measurements of $K_L-K_S$ 
interference as a function of time should represent a convenient means to 
determine the $(3\pi)$ rescattering matrix elements, because this observable 
depends linearly on ${\Im m}[(A^S_{+-0})^{*} A^L_{+-0}]$. To this purpose, 
``interferometry machines'' such as DA$\Phi$NE and LEAR should have the 
advantage that interference naturally occurs in vacuum there. It is possible 
to measure $K_L-K_S$ interference terms also in fixed-target experiments 
where statistics can be higher, however in this case an accurate knowledge of 
the regeneration amplitude is required.

Recent LEAR data \cite{nakada} give a preliminary indication of the 
term proportional to $\cos(\Delta mt)$ in Eq. (1) and suggest 
the possibility of measuring in the near future also the $\sin(\Delta mt)$ 
component. Consequently, it is worthwhile to improve the order of magnitude 
estimates  of Ref. \cite{dambrosio1} and, using the ChPT results of 
sects. \ref{sec:fsi}, \ref{sec:dalitz}, to derive predictions for 
machines like LEAR and DA$\Phi$NE based on that definite theoretical model.

Choosing ${\vert{\overline{K^0}}\rangle=CP\vert K^0\rangle}$, the 
CP-even and CP-odd eigenstates are 
$\vert K_{1,2}\rangle=(\vert K^0\rangle\pm\vert{\overline K^0}\rangle)/
\sqrt{2}$, and, with the Wu-Yang phase convention, 
the mass eigenstates are (assuming $CPT$ invariance)
\begin{equation}
\vert K_{S,L}\rangle=p\vert K^0\rangle\pm q\vert{\overline{K^0}}\rangle\equiv
\frac{\vert K_{1,2}\rangle+{\varepsilon}
\vert K_{2,1}\rangle}{\sqrt{1+\vert\varepsilon\vert^2}}.  
\label{masseigen}\end{equation}  
The proper time evolution of initial $K^0$ or ${\overline{K^0}}$ states is
\begin{eqnarray}
\vert K^0(t)\rangle=\frac{\sqrt{1+\vert\varepsilon\vert^2}}
{\sqrt 2\left(1+
{\varepsilon}\right)}\left[\vert K_S\rangle
\exp{\left(\frac{-\Gamma_St}{2}-im_St\right)}+
\vert K_L\rangle
\exp{\left(\frac{-\Gamma_Lt}{2}-im_Lt\right)}\right],\nonumber\\ 
\vert{\overline{K^0}}(t)\rangle=\frac{\sqrt{1+\vert\varepsilon\vert^2}}
{\sqrt 2\left(1-
{\varepsilon}\right)}\left[\vert K_S\rangle
\exp{\left(\frac{-\Gamma_St}{2}-im_St\right)}-
\vert K_L\rangle
\exp{\left(\frac{-\Gamma_Lt}{2}-im_Lt\right)}\right].
\label{kevolution}\end{eqnarray}

At LEAR, tagged $K^0$ and ${\overline{K^0}}$ are produced, and the simplest 
means to observe interference is represented by the asymmetry
\begin{eqnarray}
A^{+-0}_f (t)=\frac{\int d\Phi\ f (X,Y)
\left[\vert A(K^0\to\pi^+\pi^-\pi^0)\vert^2-
\vert A({\overline{K^0}}\to\pi^+\pi^-\pi^0)\vert^2\right]}
{\int d\Phi\left[\vert A(K^0\to\pi^+\pi^-\pi^0)\vert^2+
\vert A({\overline{K^0}}\to\pi^+\pi^-\pi^0)\vert^2\right]}\label{af}
\end{eqnarray}  
where $f(X,Y)$ is an odd-$X$ function  chosen in order to disentangle the 
different kinematical dependences.  Up to first order in $\varepsilon$, the 
decay amplitude squared as a function of time is given by
\begin{eqnarray}
\vert A\left(K^0({\overline{K^0}})\to \pi^+\pi^-\pi^0\right)\vert^2
\simeq\frac{1}{2}
\left(1\mp 2{\Re e}\hskip 2pt\varepsilon\right)\big\{\exp{(-\Gamma_St)}
\vert A_S\vert^2+\exp{(-\Gamma_Lt)}\vert A_L\vert^2\nonumber \\
\pm 2\exp{(-\Gamma t)}\left[{\Re e}\hskip 2pt\left(A_LA_S^*\right)
\cos{(\Delta mt)}+ 
{\Im m}\hskip 2pt\left(A_LA_S^*\right)\sin{(\Delta mt)}\right]
\big\},\label{a2evolution}\end{eqnarray}
where $\Delta m=m_L-m_S$, $\Gamma=(\Gamma_L+\Gamma_S)/2$ 
and ${A_{S,L}\equiv A^{S,L}_{+-0}}$. Then Eq. (\ref{af}) can be rewritten as 
\begin{equation}
A^{+-0}_f (t)=\frac{2e^{-\Gamma t}\int d\Phi\ f (X,Y)\hskip 2pt 
{\Re e} A_L\hskip 2pt{\Re e} A_S}{\int d\Phi\left[e^{-\Gamma_S t}
\vert A_S\vert^2+e^{-\Gamma_L t}\vert A_L\vert^2 \right]}
\left[\cos{(\Delta mt)}+\delta_f \sin{(\Delta mt)} +O(\delta_f^2) \right],
\label{asy1}\end{equation}
where
\begin{equation}
\delta_f=\frac{\int d\Phi\ f (X,Y)\left[
{\Im m} A_L\hskip 2pt{\Re e} A_S-{\Im m} A_S\hskip 2pt{\Re e} A_L\right]}
{\int d\Phi\ f (X,Y)\hskip 2pt{\Re e} A_L\hskip 2pt{\Re e} A_S}.
\label{df}\end{equation}

At the planned DA$\Phi$NE machine, a $K_S-K_L$ coherent state will be 
produced and the interference term of Eq. (\ref{int1}) 
can be studied by looking at the final state $(l^\mp \pi^\pm \nu , 
\pi^+\pi^-\pi^0)$ \cite{dambrosio1}. Following Ref. \cite{dunietz}, 
we define for a generic decay $K_{S,L}K_{L,S}\to f_1(t_1)f_2(t_2)$
an intensity
\begin{equation}
I(f_1,f_2;t)={1\over 2}\int^\infty_{\vert t \vert} dT
\vert\langle f_1(t_1)f_2(t_2)\vert i \rangle \vert^2,
\end{equation}
where $t=t_1-t_2$ and $T=t_1+t_2$. Choosing
 $f_1=l^\mp \pi^\pm \nu$ and $f_2=\pi^+\pi^-\pi^0$, we can define
an asymmetry similar to $A^{+-0}_f (t)$, namely
\begin{eqnarray}
R^\pm_f (t) = {\int d\Phi\ f (X,Y)\hskip 2pt 
I(l^\mp \pi^\pm \nu,\pi^+\pi^-\pi^0; t)
\over \int d\Phi\ I(l^\mp \pi^\pm \nu,\pi^+\pi^-\pi^0; t)}.
\end{eqnarray}
Indeed, for $t>0$ we have 
\begin{eqnarray}
I(l^\mp \pi^\pm \nu , \pi^+\pi^-\pi^0; t>0)=
 {\Gamma_L (l^\mp \pi^\pm \nu) \over 2\Gamma }\Big\{
\exp{(-\Gamma_St)}
\vert A_L\vert^2+\exp{(-\Gamma_Lt)}\vert A_S\vert^2 \nonumber \\
\pm 2\exp{(-\Gamma t)}\left[{\Re e}\hskip 2pt\left(A_LA_S^*\right)
\cos{(\Delta mt)}- {\Im m}\hskip 2pt\left(A_LA_S^*\right)\sin{(\Delta mt)}
\right] \Big\},\label{pippo}
\end{eqnarray}
and therefore
\begin{equation}
R^\pm_f (t>0)=\pm \frac{2e^{-\Gamma t}\int d\Phi\ f (X,Y)\hskip 2pt
{\Re e} A_L\hskip 2pt{\Re e} A_S}{\int d\Phi\left[e^{-\Gamma_L t}
\vert A_S\vert^2+e^{-\Gamma_S t}\vert A_L\vert^2 \right]}
\left[\cos{(\Delta mt)}-\delta_f \sin{(\Delta mt)}\right],
\label{dafne1}\end{equation}
where $\delta_f$ is the same as defined in Eq. (\ref{df}) and terms of order 
$\delta_f^2$ have been neglected. For $t<0$ the analogue of 
Eq. (\ref{dafne1}) is 
\begin{equation}
R^\pm_f (t<0)=\pm \frac{2e^{-\Gamma\vert t\vert}\int d\Phi\ f (X,Y)\hskip 2pt 
{\Re e} A_L\hskip 2pt{\Re e} A_S}{\int d\Phi\left[e^{-\Gamma_S\vert t\vert}
\vert A_S\vert^2+e^{-\Gamma_L \vert t\vert}\vert A_L\vert^2 \right]}
\left[\cos{(\Delta m\vert t\vert)}+\delta_f \sin{(\Delta m\vert t\vert)} 
\right].\label{dafne2}\end{equation}
However, due to the exchange $\Gamma_L\leftrightarrow\Gamma_S$, the 
denominator in Eq. (\ref{dafne2}) quickly becomes much larger than in 
Eq. (\ref{dafne1}), and suppresses the interference effect. 

Considering for $\delta_f$ the first non-vanishing order in ChPT, which is 
$O(p^6)$ in the numerator and $O(p^4)$ in the denominator, we obtain 
\begin{equation}
\delta_f=\frac{\int d\Phi\ f (X,Y)\left[
a_n(\alpha-\alpha'-{\widetilde \delta})X-b_n(\beta-\beta'-
{\widetilde \delta})XY\right]}
{\int d\Phi\ f (X,Y)\left[ a_nX-b_nXY\right]}
\label{df2}\end{equation}
If we use as weight function $f(X,Y)={\rm sgn}(X)$, we obtain numerically
the following result:
\begin{equation}
\delta_X=0.18\pm 0.01\hskip 2pt .
\label{dx}\end{equation}
Essentially, this turns out to be: $\delta_X=\alpha_0-\delta_0$, and 
is practically independent of the theoretical uncertainties on the small 
ratio $b_n/a_n$. For this reason the result (\ref{dx}) is in 
good agreement with the prediction of Ref. \cite{dambrosio1}. 

On the other hand, choosing $f(X,Y)={\rm sgn}(YX)$ we obtain numerically: 
\begin{equation}
\delta_{XY}=0.30\pm 0.05.
\label{dxy}\end{equation}
This result is about a factor four larger than obtained in \cite{dambrosio1}, 
$\delta_{XY}\simeq 0.07$. Indeed, by expanding the rescattering functions, 
in the present calculation we have  
\begin{equation}
\delta_{XY}\simeq \frac{(\beta_0-\delta_0)+
{\displaystyle a_n \over \displaystyle b_n}(\alpha'_0-2\delta_1)}
{1- {\displaystyle a_n \over \displaystyle b_n}
\frac{\displaystyle\int d\Phi\ \vert X
\vert {\rm sgn}(Y)}{\displaystyle \int d\Phi\ \vert XY\vert }}.
\label{delxy}\end{equation}
Eq. (\ref{delxy}) shows that also in the case of $\delta_{XY}$ the 
Y-dependent terms give a sizable contribution, since they are multiplied 
by the large factor $(a_n/b_n)$. 
The error in Eq. (\ref{dxy}) accounts for the theoretical uncertainty on 
$(a_n/b_n)$, for which either the experimental value or the O($p^2$)
ChPT prediction can be used.\footnote{The unknown quadratic term in $A^S_{+-0}$ 
can affect to some extent the numerical result for 
$\delta_{XY}$ \cite{kambor}, and its effect can be roughly taken into account 
by doubling the error in Eq. (\ref{dxy}).}

\section{Concluding remarks}
\label{sec:concl}
In the previous sections we have introduced a general formalism to 
consistently account for final states 
interactions in ${K\rightarrow 3\pi}$ amplitudes, 
and have used leading order Chiral Perturbation Theory to evaluate the 
rescattering matrix. We have considered some potentially observable effects 
of rescattering on Dalitz plot variables. The results indicate that the 
off-diagonal elements of the rescattering matrix in the $I=1$ sector 
induce sizable imaginary parts in the $X$ and $Y$ dependent amplitudes. 
However, these large imaginary parts are not easily detected from
Dalitz plot analyses and cancel in direct CP-violating asymmetries.

Planned experiments at ``interferometry machines'' can have direct access 
to the rescattering matrix elements {\it via} appropriately defined 
time-dependent asymmetries, which we have estimated 
in leading order ChPT. As examples of the typical effects expected in this 
framework,  Fig. 2 shows the asymmetry $A^{+-0}_X (t)$ of Eq. (\ref{asy1}) 
relevant to LEAR. The solid line represents the asymmetry with no rescattering 
($\delta_X=0$), the dashed line corresponds to the leading ChPT estimate 
of Eq. (\ref{dx}), and finally the dotted line would result by 
doubling the value of $\delta_X$. To obtain Fig. 2, for the real parts of 
the amplitudes $A^L_{+-0}$ and $A^S_{+-0}$ we have used the expansion 
(\ref{expan11}) with the values of the parameters obtained in the fit of 
\cite{kambor}. As one can see, the curves in this figure have similar shapes, 
but possibly could be distinguished in high precision experiments. 

In Fig. 3 we report the asymmetry $R^+_X(t)$ of Eq. (\ref{dafne1}) relevant 
to DA$\Phi$NE, and the three curves refer to the same cases considered 
in Fig. 2. Here we note that rescattering affects the shape of the curves 
more significantly, especially for $t>0$ where  the asymmetry can become 
quite large. However, this occurs for the values of $t$ where the number of 
events becomes smaller. As an indication, the total expected number of 
events at $t>0$, with the planned DA$\Phi$NE luminosity 
$5\times 10^{32} cm^{-2}sec^{-1}$, is of the order of $10^3$/year. 

In conclusion, the analysis presented here shows the interest of experimental 
efforts to accurately measure the kind of asymmetries discussed here. 
The ultimate goal would be the determination of 
the ${K\rightarrow 3\pi}$ rescattering 
matrix elements testing ChPT in the strong sector, but in any case even a 
reasonable upper bound would represent an important information in this 
regard. Furthermore, the direct measurement of the CP conserving 
$K_S\to 3\pi$ amplitude is by itself an important achievement, extremely 
useful in order to test chiral symmetry in non-leptonic weak interactions. 

\acknowledgements
We would like to thank L. Maiani for useful discussions.

\begin{figure}
\bigskip
\caption{Loop diagrams relevant to ${K\rightarrow 3\pi}$ rescattering. 
The symbols 
$\protect{\bullet}$ 
and $\protect{\circ}$ indicate the weak and the strong vertices, respectively.}
\bigskip
\caption{The asymmetry $A_X^{+-0}$ of Eq. (\protect{\ref{asy1}}) vs. 
$t$. The full, dashed and dotted lines correspond to 
$\protect{\delta_X}=0$, 0.2 and 0.4, respectively.}
\bigskip
\caption{The asymmetry $R_X^{+}$ of Eq. (\protect{\ref{dafne1}}) 
for positive and negative $t$. The full, dashed and dotted lines 
correspond to $\protect{\delta_X}=0$, 0.2 and 0.4, respectively.}
\end{figure}

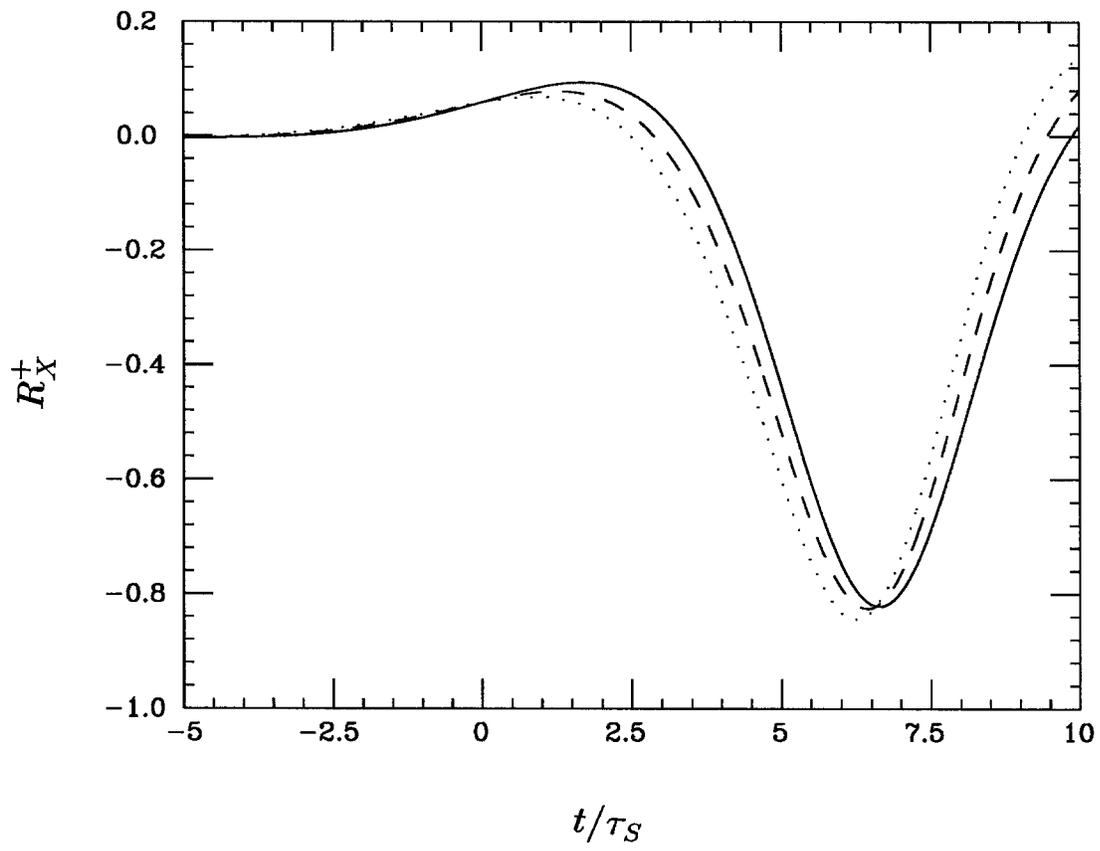
\end{document}